\documentclass[twocolumn]{aastex631}
\shorttitle{Activity patterns of V711~Tau}
\shortauthors{Cao \& Gu}
\graphicspath{{./}{figures/}}

\begin{document}

\title{Further Investigation on Chromospheric Activity Patterns of the RS Canum Venaticorum Star V711~Tauri}

\correspondingauthor{Dongtao Cao, Shenghong Gu}
\email{dtcao@ynao.ac.cn, shenghonggu@ynao.ac.cn}

\author[0000-0002-3534-1740]{Dongtao Cao}
\affiliation{Yunnan Observatories, Chinese Academy of Sciences, Kunming 650216, China}
\affiliation{Key Laboratory for the Structure and Evolution of Celestial Objects, Chinese Academy of Sciences, Kunming 650216, China}
\affiliation{International Centre of Supernovae, Yunnan Key Laboratory, Kunming 650216, China}

\author{Shenghong Gu}
\affiliation{Yunnan Observatories, Chinese Academy of Sciences, Kunming 650216, China}
\affiliation{Key Laboratory for the Structure and Evolution of Celestial Objects, Chinese Academy of Sciences, Kunming 650216, China}
\affiliation{School of Astronomy and Space Science, University of Chinese Academy of Sciences, Beijing 101408, China}



\begin{abstract}
We present new long-term spectroscopic observations of the very active RS Canum Venaticorum (RS CVn) star V711~Tauri (V711~Tau), obtained during several observing runs from 2005 to 2021. By means of spectral subtraction technique, several optical chromospheric activity indicators $\mbox{Ca~{\sc ii}}$ infrared triplet, H$_{\alpha}$, $\mbox{Na~{\sc i}}$ D$_{1}$, D$_{2}$ doublet, $\mbox{He~{\sc i}}$ D$_{3}$, and H$_{\beta}$ lines, are analyzed. V711~Tau shows strong excess emission in the chromospheric activity lines, and the emission is mainly associated with the K1~IV primary component. The G5~V secondary component also shows weak emission features. There are four optical flare events during our observations, characterized by the $\mbox{He~{\sc i}}$ D$_{3}$ line emission features and stronger emission in other activity lines, which have similar order of magnitude as the strong flares of other active RS CVn-type stars. We separate our long-term observations to twelve observing groups, and find that V711~Tau has two active longitudes around phases 0.25 and 0.75 to dominate the activity in most of time. The chromospheric activity level shows a long-term variation, indicating possible chromospheric activity cycle. Moreover, a close spatial connection between photospheric spots and chromospheric active regions in short and long timescale is found for V711~Tau.
\end{abstract} 

\keywords{Stellar activity (1580) --- Optical ﬂares (1166) --- Stellar chromospheres (230) --- Spectroscopy (1558) --- Spectroscopic binary stars (1557)}

\section{Introduction} \label{sec1}
RS~CVn-type stars are well known due to their strong magnetic activity in several forms, such as photospheric starspots, chromospheric plages, and coronal radiation, as well as intense flares in the optical, ultraviolet (UV), and X-ray wavelength. It is commonly accepted that all of these active phenomena arise from a powerful magnetic dynamo generated by the interplay between the turbulent motions in the convection zone and the stellar differential rotation, in a manner similar to the solar case. Moreover, having fast rotation rates induced by tidal synchronization, RS~CVn-type stars usually display a high level of activity with strong chromospheric line emissions, and are more active than single stars of similar mass and age. Since 1997, at Yunnan Observatories, we have begun a long-term high-resolution spectroscopic monitoring project for a number of RS CVn-type systems, to study their magnetic activity behavior (detecting optical flares, searching for prominence-like events, exploring the rotational modulation of chromospheric activity, and investigating long-term variations of chromospheric activity and possible activity cycles) based on the information derived through several optical chromospheric activity indicators \citep[e.g.][]{gu2002, zhang2008, cao2012, cao2015, cao2017, cao2019, cao2020}. In our previous paper (\citealt{cao2015}, hereafter C15), we had focused on one of the most bright and active RS CVn-type binaries, V711~Tau. Long-term high-resolution spectroscopic observations, taken during several observing runs from 1998 to 2004, were analyzed to study its chromospheric activity. The result shows that V711~Tau possesses various and strong solar-type activity phenomena.

V711~Tau (=~HR 1099~=~HD 22468) is one of the most extensively studied members of the RS~CVn-type stars. It is a close double-lined, non-eclipsing spectroscopic binary system consisting of a K1~IV primary component and a G5~V companion in an almost circular orbit with a period of about 2.84~days \citep{fekel1983, donati1999}. Strong chromospheric activity of V711~Tau could be demonstrated by the H$_{\alpha}$ line emission above the continuum at all times, similar to very active RS CVn-type stars II~Peg, UX~Ari, and DM~UMa. Its chromospheric activity has been studied by several authors, e.g. \citet{gondoin1986}, \citet{montes1995b}, \citet{zhai1996}, \citet{montes1997}, and \citet{garcia2003a}. It is well accepted that V711~Tau shows the pronounced chromospheric emission in several activity lines, such as $\mbox{Ca~{\sc ii}}$ H and K, H$_{\alpha}$, and $\mbox{Ca~{\sc ii}}$ IRT lines, and the emission features are mainly associated with the K1~IV primary star. Moreover, based on Doppler imaging technique, several authors, e.g. \citet{vogt1999}, \citet{donati1999}, \citet{strassmeier2000}, \citet{garcia2003b}, \citet{donati2003}, and \citet{petit2004}, have studied the starspot distribution on the surface of V711~Tau. Using the Zeeman Doppler imaging technique, \citet{donati1999} also detected strong magnetic field of the order of 1000~G on V711~Tau.

Generally, most of long-term stellar activity studies of RS CVn-type stars are based on easily detected optical photometric variations caused by starspots. V711~Tau shows long-term starspot evolution varied cyclically within several years \citep{henry1995, lanza2006, berdyugina2007, muneer2010, Jetsu2017}. Moreover, magnetic activity is also periodic in the other wavelength bands. For example,  \citet{buccino2009} analyzed possible chromospheric activity cycle based on $International$ $Ultraviolet$ $Explorer$ $(IUE)$ UV high and low-resolution $\mbox{Mg~{\sc ii}}$ h and k lines from 1975 to 1996. Long-term variations in the X-ray activity of V711~Tau, based on archival data from 1978 to 2017, had been investigated by \citet{Perdelwitz2018}. Due to sparse data sampling and stochastic flaring activity, however, it is not allowed to independently determine of an X-ray activity cycle.

To understand the structure of active regions in the atmosphere and their evolution in detail, long-term observations are necessary. To further investigate the chromospheric activity of V711~Tau, we present more new high-resolution spectroscopic observations taken from 2005 to 2021, which cover several optical chromospheric activity indicators formed at different atmospheric heights. The details of new long-term high-resolution spectroscopic observations and data reduction are given in Section~\ref{sec2}, and the procedure of the spectral analysis and results are described in Section~\ref{sec3}. In Section~\ref{sec4}, optical flare events detected during our observations are discussed. The rotational modulation and long-term variations of chromospheric activity, and the correlation between photospheric spots and chromospheric activity regions are investigated in Section~\ref{sec5}. Finally, we conclude and summarize our new results in Section~\ref{sec6}.

\section{Spectroscopic observations and data reduction}\label{sec2}
Long-term high-resolution spectroscopic observations of V711~Tau analyzed here were obtained during some observing runs from 2005 to 2021. During the runs from 2005 to 2008, the observations were carried out with the coud\'{e} echelle spectrograph (CES, \citealt{zhao2001}) mounted on the 2.16~m telescope at the Xinglong station, National Astronomical Observatories, Chinese Academy of Sciences, China. The echelle spectra were recorded on a $1024 \times 1024$-pixels Tektronix CCD detector, which cover the wavelength range of about 5600--9000~\AA~with an average resolving power of R~=~$\lambda$/$\Delta\lambda$~$\simeq$~37000. Several main chromospheric activity indicators, including $\mbox{Ca~{\sc ii}}$ IRT, H$_{\alpha}$, $\mbox{Na~{\sc i}}$ D$_{1}$, D$_{2}$ doublet, and $\mbox{He~{\sc i}}$ D$_{3}$ lines, are covered in the spectra. Moreover, a fiber-fed high-resolution spectrograph (HRS) with a spectral resolving power of R~=~$\lambda$/$\Delta\lambda$~$\simeq$~48000 over the wavelength range of about 3900--10000~\AA~and a $4096 \times 4096$ pixels EEV CCD detector, installed on the 2.16~m telescope later, were used to obtain spectra during the observing runs from 2015 to 2021, except the observations of 2017 January~22. During this night, the observations were taken from the 2.4~m telescope \citep{Fan2015} at the Lijiang station of Yunnan observatories, Chinese Academy of Sciences, China, which also has a same HRS instrument \citep{Wang2019}. Because the signal to noise ratio (S/N) is very low in the $\mbox{Ca~{\sc ii}}$ H \& K line regions for the HRS observations, the spectra mainly include chromospheric activity indicators $\mbox{Ca~{\sc ii}}$ IRT, H$_{\alpha}$, $\mbox{Na~{\sc i}}$ D$_{1}$, D$_{2}$ doublet, $\mbox{He~{\sc i}}$ D$_{3}$, and H$_{\beta}$ lines.

We give a detailed observing log in Table~\ref{tab1}, which includes the observing date, the heliocentric Julian date (HJD), orbital phase, and exposure time. The orbital phases were calculated with the ephemeris:
\begin{equation}
T_{0}(HJD)=2,457,729.7084+2^{d}.837711~\times~E,
\end{equation}
from \citet{Strassmeier2020}, where the epoch corresponds to the K1~IV primary star of V711~Tau with maximum positive radial velocity. Because of different ephmeris used, there is an offset of 0.25 between the orbital phases here and the ones calculated in C15. In addition, besides V711~Tau, observations of some rapidly rotating early-type stars and slowly rotating inactive stars with the same spectral type and luminosity class as each component of the system were also obtained for each observing run. The spectra of early-type stars were used as telluric templates whereas the inactive stars were used as references in the spectral subtraction technique.

The data reduction was performed with the IRAF\footnote{IRAF is distributed by the National Optical Astronomy Observatories, which is operated by the Association of Universities for Research in Astronomy (AURA), Inc., under cooperative agreement with the National Science Foundation.} package, following the prescription in \citet{Cao2024}. The wavelength calibration was obtained by using a lot of emission lines of a Th-Ar lamp, and all spectra were normalized using low-order polynomial fit to the observed continuum. Finally, for some of our observations, there are heavy telluric absorption lines in the chromospheric activity line regions of interest, which mainly include $\mbox{Ca~{\sc ii}}$~$\lambda$8498, H$_{\alpha}$, $\mbox{Na~{\sc i}}$ D$_{1}$, D$_{2}$ doublet, and $\mbox{He~{\sc i}}$ D$_{3}$ line regions. We eliminated them using the spectra of four brighter and rapidly rotating early-type stars HR~8858 (B5~V, $vsini$ = 316~km~s$^{-1}$), HR~7894 (B5~IV, $vsini$ = 285~km~s$^{-1}$), HR~989 (B5~V, $vsini$ = 260~km~s$^{-1}$) and HR~1051 (B8~V, $vsini$ = 295~km~s$^{-1}$), respectively, with an interactive procedure in the IRAF package described in detail by \citet{gu2002}.

\begin{figure*}
\includegraphics[width=16.cm,height=16.cm]{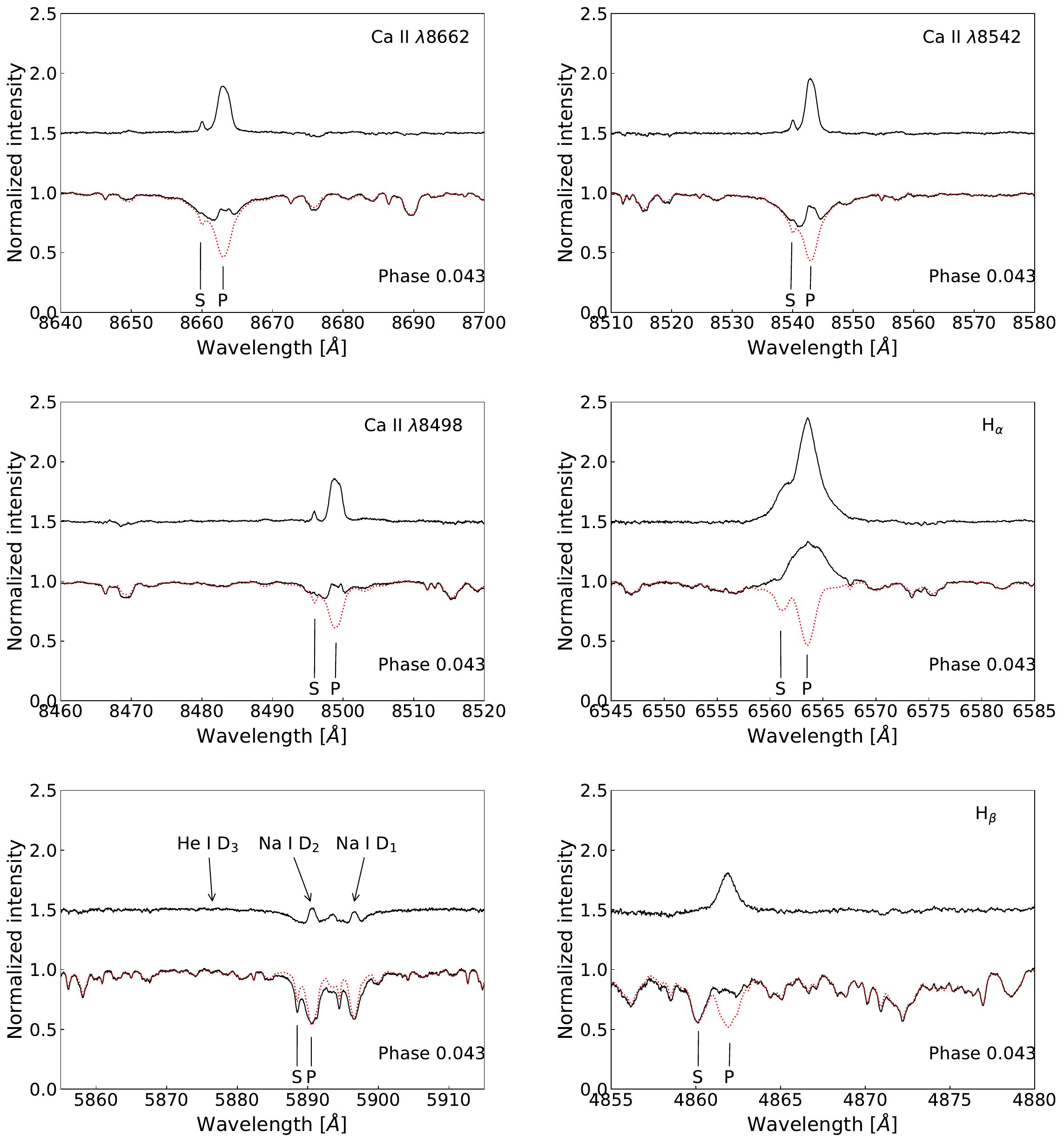}
\centering
\caption{Examples of the observed, synthesized, and subtracted spectra for the $\mbox{Ca~{\sc ii}}$~$\lambda$8662, $\mbox{Ca~{\sc ii}}$~$\lambda$8542, and $\mbox{Ca~{\sc ii}}$~$\lambda$8498, H$_{\alpha}$, $\mbox{Na~{\sc i}}$ D$_{1}$, D$_{2}$ doublet, $\mbox{He~{\sc i}}$ D$_{3}$, and H$_{\beta}$ line regions. For each panel, the lower solid-line is the observed spectrum, the dotted red line represents the synthesized spectrum and the upper spectrum is the subtracted one, shifted for better display. ``P'' and ``S" indicate the primary and secondary components of V711~Tau, respectively. The label identifying each chromospheric activity indicator is marked in the corresponding panel. Here, regarding the prominent absorption feature labeled ``S" in the H$_{\beta}$ line region, in addition to the contribution from the H$_{\beta}$ line of the G5~V secondary star, there is also a notable contribution from an absorption line of the K1~IV primary star, presumably the $\mbox{Fe~{\sc i}}$~$\lambda$4859.7 line. These two contributions combined to from a prominent absorption feature.}
\label{Fig1}
\end{figure*}

In Figure~\ref{Fig1}, we display examples of the normalized $\mbox{Ca~{\sc ii}}$ IRT, H$_{\alpha}$, $\mbox{Na~{\sc i}}$ D$_{1}$, D$_{2}$ doublet, $\mbox{He~{\sc i}}$ D$_{3}$, and H$_{\beta}$ line regions of V711~Tau obtained by the HRS instrument of the Xinglong 2.16~m telescope at phase 0.043 on 2016 November~14.

\begin{deluxetable}{cccc}
\tablenum{1}
\tablecaption{Observing Log of V711~Tau\label{tab1}}
\tablewidth{0pt}
\tablehead{
\colhead{\bf{Date}} &\colhead{\bf{HJD}} &\colhead{\bf{Phase}} &\colhead{\bf{Exp.time}}\\
\nocolhead{} & \colhead{(2,450,000+)} & & \colhead{(s)}
}
\startdata
2005-11-18 & 3693.23949 & 0.562 & 1800\\ 
2005-11-18 & 3693.26108 & 0.569 & 1800\\ 
2005-11-20 & 3695.29036 & 0.284 & 1800\\ 
2005-11-20 & 3695.31218 & 0.292 & 1800\\ 
2005-11-21 & 3696.31727 & 0.646 & 1200\\ 
2005-11-21 & 3696.33135 & 0.651 & 1200\\ 
\enddata
\tablecomments{Table~\ref{tab1} is published in its entirety in the machine-readable format. A portion is shown here for guidance regarding its form and content.}
\end{deluxetable}
\begin{figure*}
\centering
\includegraphics[width=18cm,height=22cm]{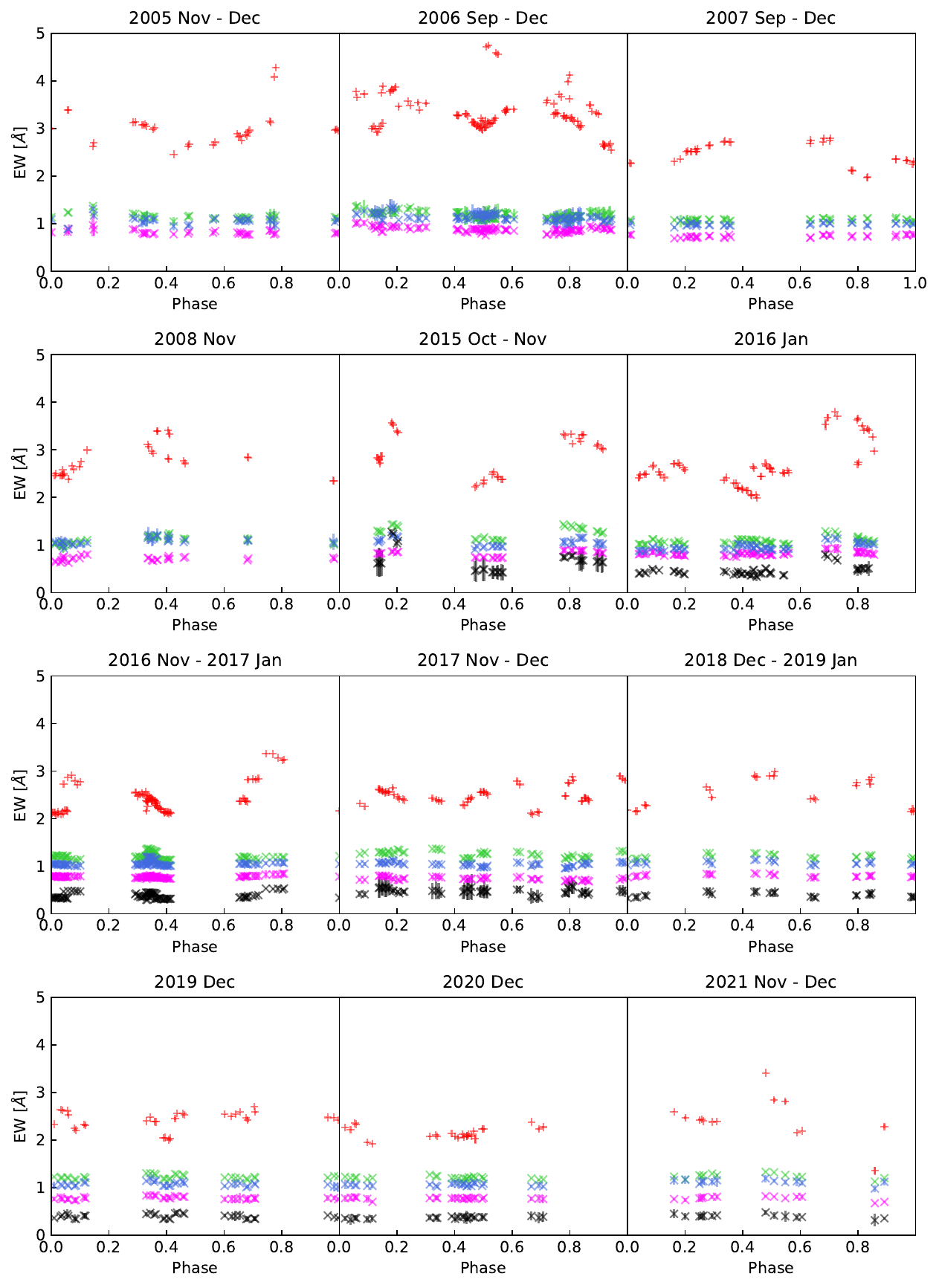}
\caption{EWs of the $\mbox{Ca~{\sc ii}}$~$\lambda$8662 (royalblue symbol), $\mbox{Ca~{\sc ii}}$~$\lambda$8542 (limegreen symbol), $\mbox{Ca~{\sc ii}}$~$\lambda$8498 (magenta symbol), H$_{\alpha}$ (red symbol), and H$_{\beta}$ (black symbol) excess emission profiles versus orbital phase.}
\label{Fig2}
\end{figure*}
\begin{figure*}
\centering
\includegraphics[width=18cm,height=22cm]{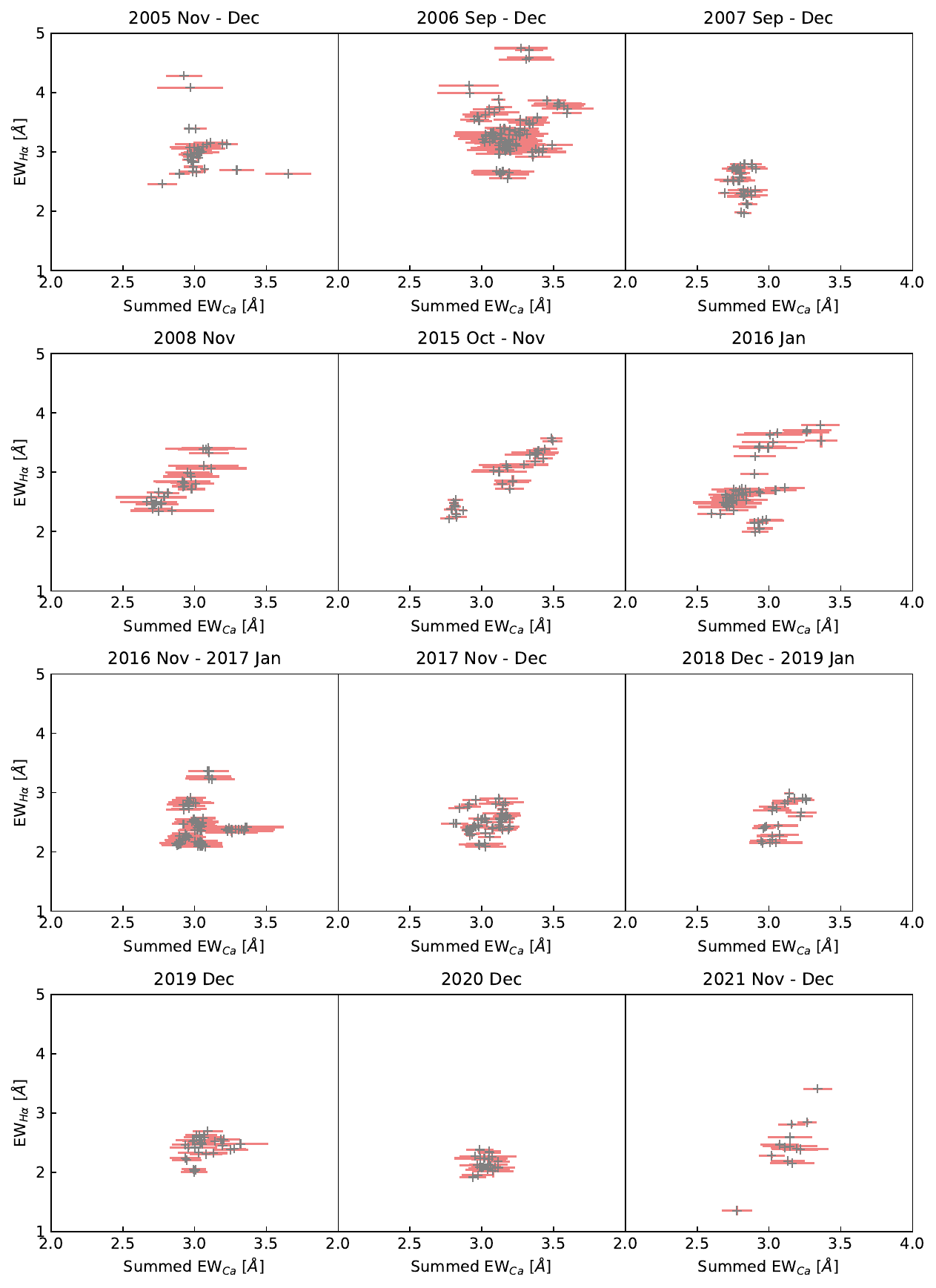}
\caption{EWs of the H$_{\alpha}$ excess emission versus summed EWs of the $\mbox{Ca~{\sc ii}}$ IRT excess emissions.}
\label{Fig2b}
\end{figure*}
\begin{figure*}
\centering
\includegraphics[width=8.75cm,height=14cm]{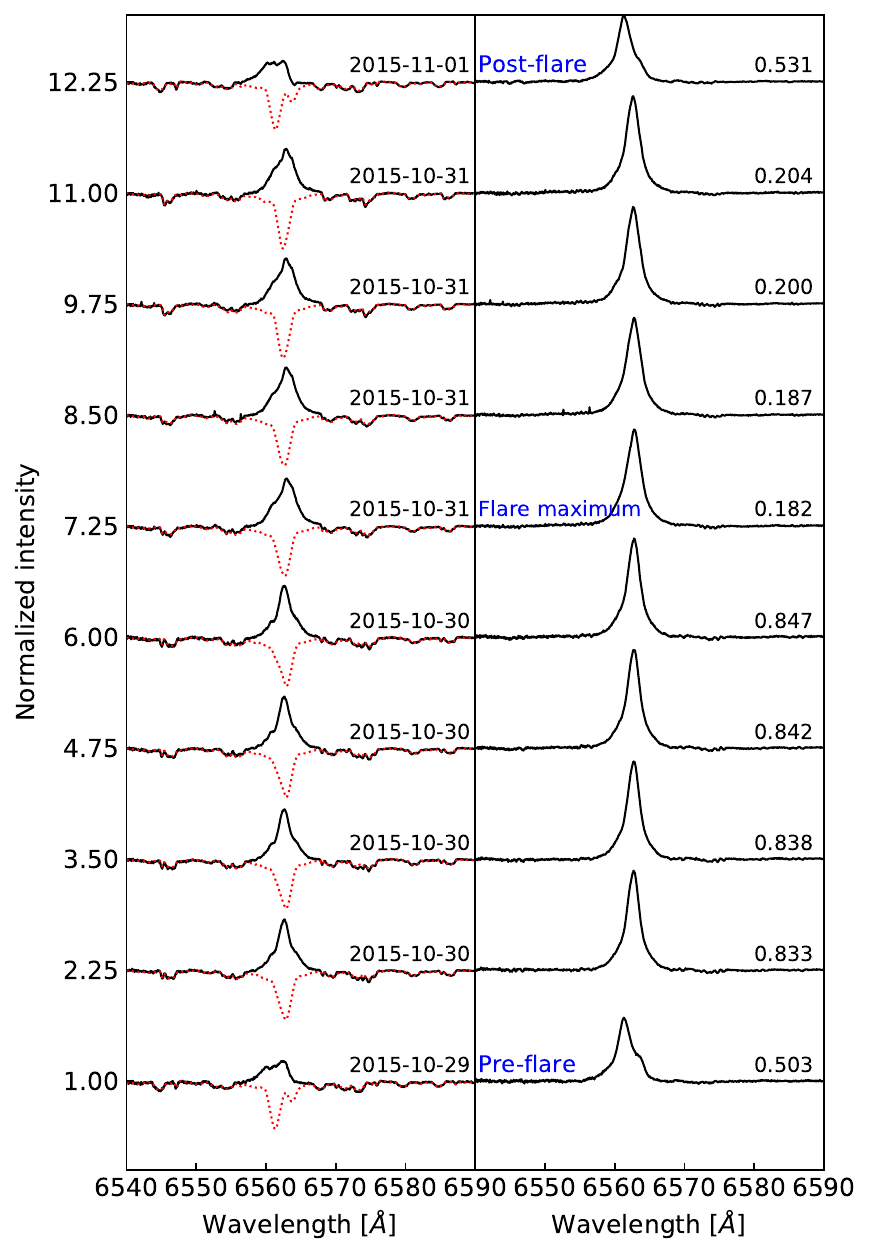}
\includegraphics[width=8.75cm,height=14cm]{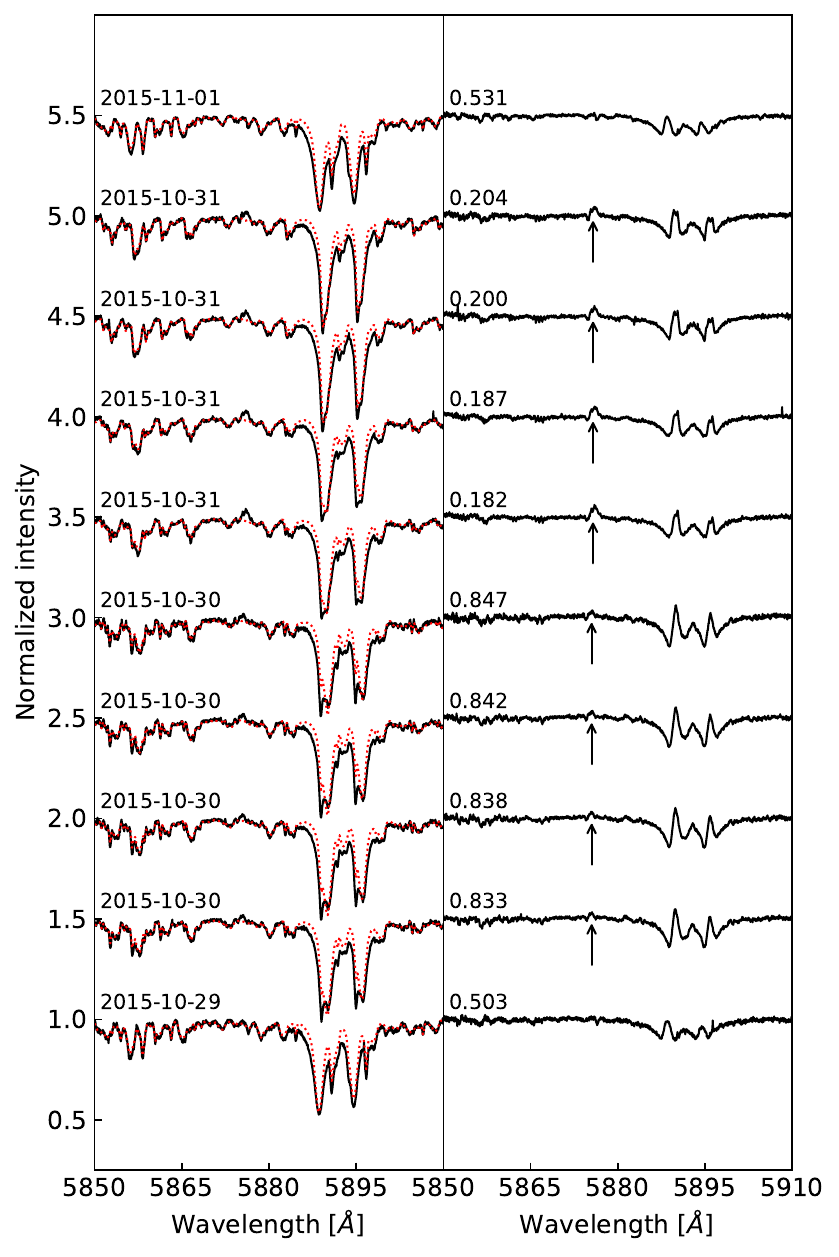}
\caption{H$_{\alpha}$, $\mbox{Na~{\sc i}}$ D$_{1}$, D$_{2}$ doublet, and $\mbox{He~{\sc i}}$ D$_{3}$ line profiles obtained during the pre-flare (2015 Oct~29), flare (2015 Oct~30 and 31), and post-flare (2015 Nov~1) periods. The observed spectra (solid line) and the synthesized ones (dotted line) are plotted in the left part of panel and the subtracted spectra in the right part. The orbital phase and observing date are also marked in each panel. Arrows indicate the $\mbox{He~{\sc i}}$ D$_{3}$ emission features in the subtracted spectra.}
\label{Figb3}
\end{figure*}
\begin{deluxetable*}{ccccc}
\tablenum{2}
\tablecaption{Optical Flare Events on V711~Tau\label{tab2}}
\tablewidth{7.5pt}
\tablehead{
\colhead{\bf{Date}} &\colhead{\bf{State}} &\colhead{\bf{Phase}}&\colhead{\bf{F$_{s}$(H$_{\alpha}$)}} &\colhead{\bf{L(H$_{\alpha}$)}}\\
\nocolhead{}&\nocolhead{}&\nocolhead{}&\colhead{($\times~10^{7}$~erg~cm$^{-2}$~s$^{-1}$)} &\colhead{($\times~10^{31}$~erg~s$^{-1}$)}
}
\startdata
\multicolumn{5}{c}{First Flare} \\
2006-11-28&Pre-flare&0.745&1.57&1.45\\
2006-11-29&Flare maximum&0.058&1.71&1.58\\
2006-12-01&Post-flare&0.762&1.67&1.55\\
&&&&\\
\multicolumn{5}{c}{Second Flare}\\
2006-12-06&Flare maximum&0.518&2.15&1.99\\
2006-12-07&Post-flare&0.870&1.57&1.46\\
&&&&\\
\multicolumn{5}{c}{Third Flare}\\
2015-10-29&Pre-flare&0.503&1.06&0.98\\
2015-10-31&Flare maximum&0.182&1.60&1.48\\
2015-11-01&Post-flare&0.531&1.11&1.03\\
&&&&\\
\multicolumn{5}{c}{Fourth Flare}\\
2016-01-22&Pre-flare&0.378&1.06&0.98\\
2016-01-23&Flare maximum&0.721&1.71&1.58\\
2016-01-24&Post-flare&0.040&1.14&1.06\\
\enddata
\end{deluxetable*}
\begin{figure*}
\centering
\includegraphics[width=18cm,height=14cm]{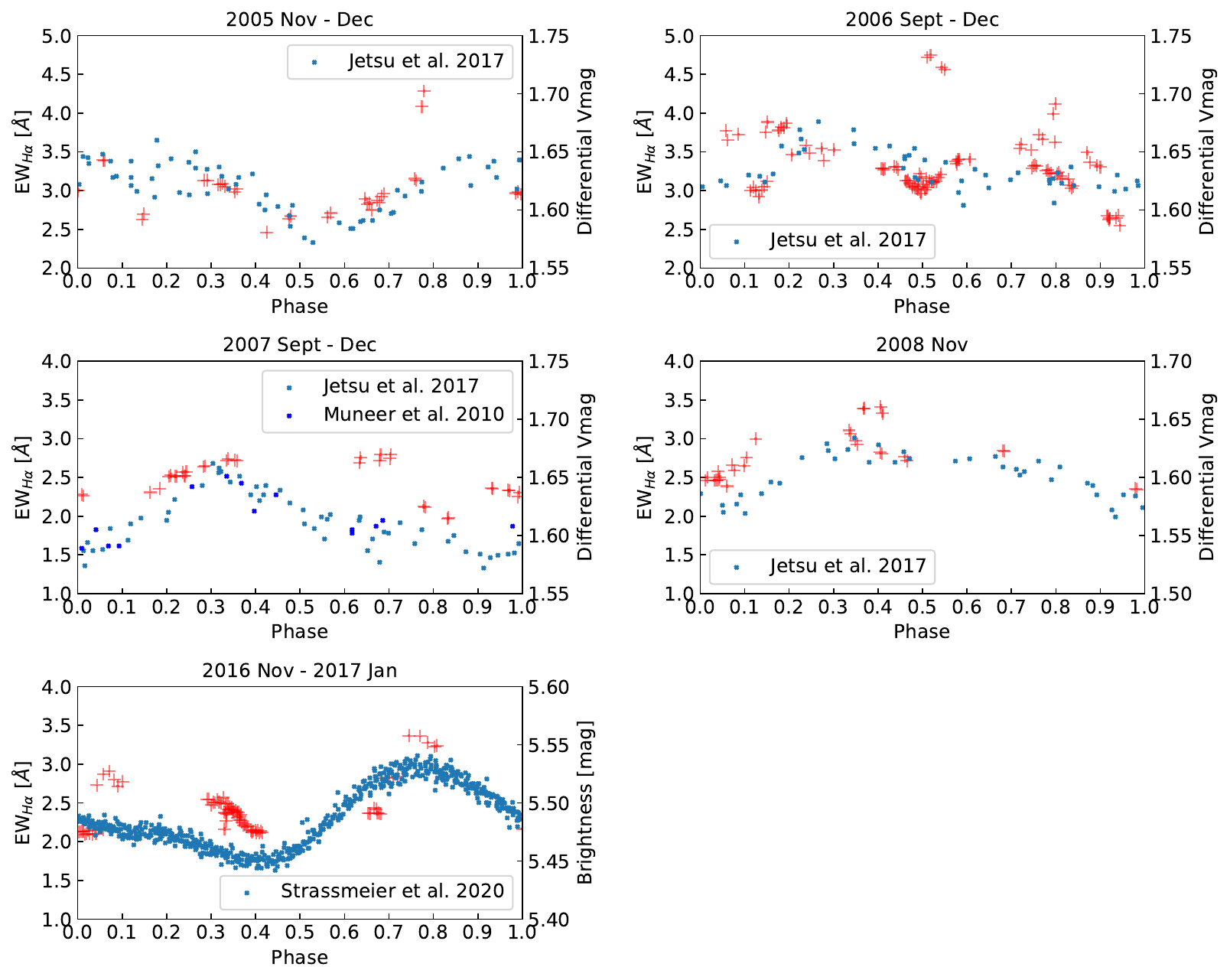}
\caption{Comparison between photospheric light curves (*) and chromospherc H$_{\alpha}$ variations (+) for the 2005 November--December, 2006 September--December, 2007 September--December, 2008 November, and 2016 November--2017 January observing groups. Phase folded light curves are taken from \citet{muneer2010}, \citet{Jetsu2017} and \citet{Strassmeier2020}, respectively.}
\label{Fig5}
\end{figure*}
\begin{figure*}
\centering
\includegraphics[width=15cm,height=9cm]{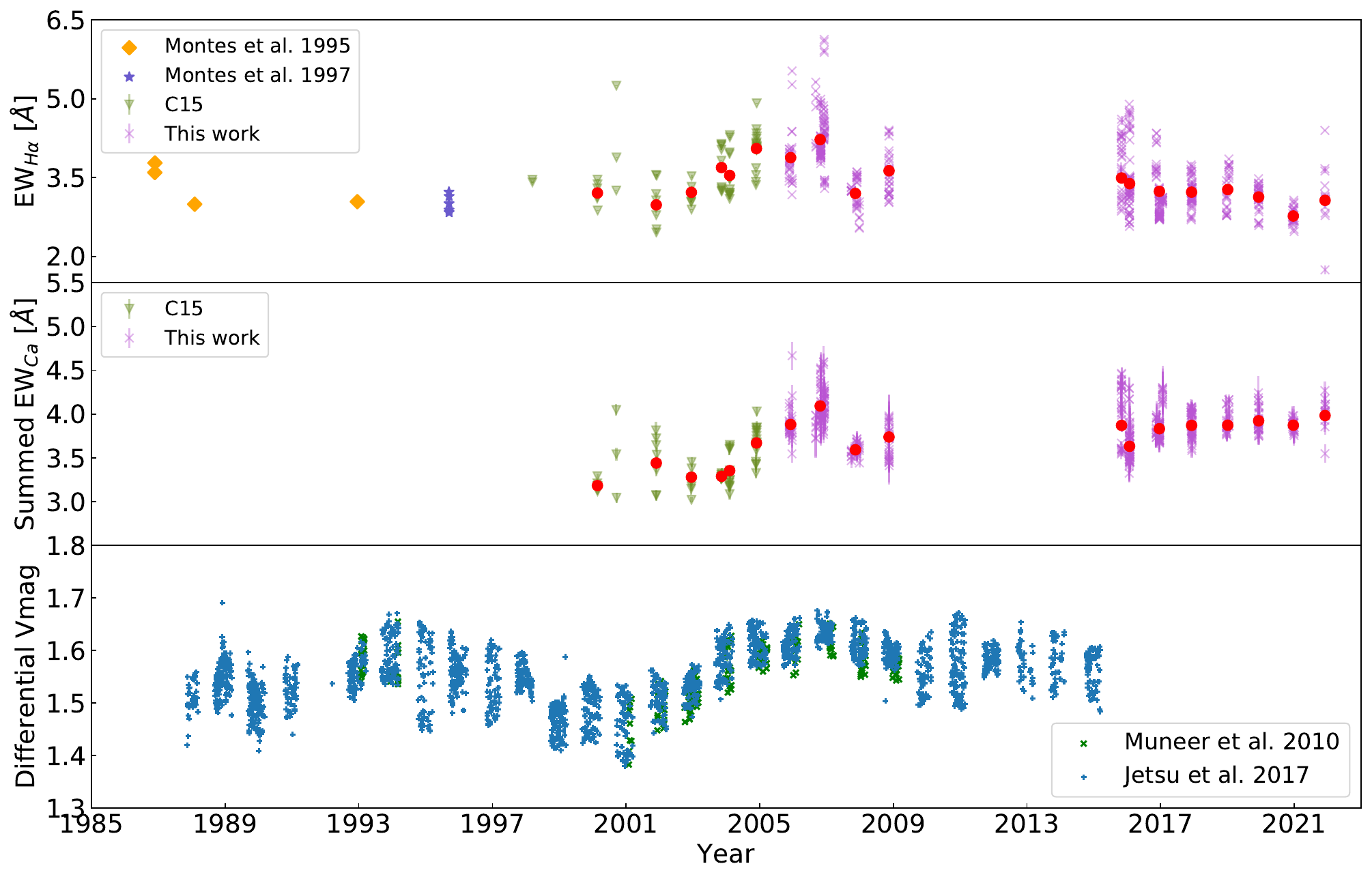}
\caption{Comparison between chromospherc emissions and photospheric starspot activities. Top and middle panels: the H$_{\alpha}$ and summed $\mbox{Ca~{\sc ii}}$ IRT excess emissions of V711~Tau versus the observing time. The data from 1986 to 2004 are taken from \citet{montes1995a}, \citet{montes1997}, and C15, respectively. The red dot represents the averaged value for each observing run. Bottom panel: photometric light curves taken from \citet{muneer2010} and \citet{Jetsu2017}.}
\label{Fig4}
\end{figure*}
\section{Spectral analysis and results}\label{sec3}
Chromospheric activity indicators $\mbox{Ca~{\sc ii}}$ IRT, H$_{\alpha}$, $\mbox{Na~{\sc i}}$ D$_{1}$, D$_{2}$ doublet, $\mbox{He~{\sc i}}$ D$_{3}$, and H$_{\beta}$ lines, formed at different atmospheric heights, are simultaneously analyzed in the present work. As shown in Figure~\ref{Fig1}, clear central emission features appear in the cores of the $\mbox{Ca~{\sc ii}}$ IRT absorption line profiles and the H$_{\alpha}$ line is in emission status above the continuum. Another Balmer line, H$_{\beta}$, frequently shows filled-in absorption profile and sometimes even appears emission above the continuum level. The $\mbox{Na~{\sc i}}$ D$_{1}$, D$_{2}$ doublet lines are characterized by deep absorption profiles. Moreover, the $\mbox{He~{\sc i}}$ D$_{3}$ line exhibits emission feature at some phases, which means that there might be optical flare events during our observations.

To isolate the chromospheric activity contribution from the observed profiles of these activity lines, we use the spectral subtraction technique for all the observed spectra with the aid of the STARMOD program \citep{barden1985, montes1997, montes2000}, which has been widely used for studies in chromospheric activity binaries. In our analysis, we use the spectra of two stars HR~3351 (K0~IV) and HR~3309 (G5~V) as references for the primary and secondary stars of V711~Tau to construct the synthesized spectra of the system. Using the same method as C15, the $vsini$ values of 41 km~s$^{-1}$ for the primary and 9.5 km~s$^{-1}$ for the secondary are obtained, and the intensity weight ratios of two components are 0.7/0.3 for the H$_{\beta}$ spectral region, 0.735/0.265 for the $\mbox{Na~{\sc i}}$ D$_{1}$, D$_{2}$, $\mbox{He~{\sc i}}$ D$_{3}$ spectral region, 0.775/0.225 for the H$_{\alpha}$ spectral region, 0.78/0.22 for the $\mbox{Ca~{\sc ii}}$ $\lambda$8498 spectral region, 0.782/0.218 for the $\lambda$8542 spectral region, and 0.785/0.215 for the $\mbox{Ca~{\sc ii}}$ $\lambda$8662 spectral region. Consequently, the synthesized spectra are constructed by broadening and weighting the reference spectra to the values of the $vsini$ and the intensity weight ratios mentioned above, and shifting along radial-velocity axis. Finally, the subtracted spectra are calculated for each chromospheric activity indicator of V711~Tau. For the CES observations, the H$_{\beta}$ line is not covered in spectra. Moreover, for some of the HRS observations in 2016 January, the H$_{\beta}$ line is not analyzed because of low S/Ns.

Examples of spectral subtraction in the $\mbox{Ca~{\sc ii}}$ IRT, H$_{\alpha}$, $\mbox{Na~{\sc i}}$ D$_{1}$, D$_{2}$ doublet, $\mbox{He~{\sc i}}$ D$_{3}$, and H$_{\beta}$ spectral regions are presented in Figure~\ref{Fig1}. The synthesized spectra match the observational ones quite well, except the $\mbox{Na~{\sc i}}$ D$_{1}$, D$_{2}$ doublet lines. It is a very common phenomenon in the synthesized spectral construction, because the $\mbox{Na~{\sc i}}$ D$_{1}$, D$_{2}$ doublet lines are very sensitive to the effective temperature, just a slight temperature difference between the target and reference stars would produce significant changes in the wings of the line profiles.

After the application of the spectral subtraction technique, V711~Tau shows strong chromospheric emission in the $\mbox{Ca~{\sc ii}}$ IRT, H$_{\alpha}$, and H$_{\beta}$ lines, consistent with the previous findings. The chromospheric emission is mainly associated with the K1~IV star, which means that the chromospheric activity is attributed to the primary star of the system. However, it is obvious that the G5~V secondary star also shows chromospheric emission features, especially at phases where the two components of V711~Tau are well separated (see the $\mbox{Ca~{\sc ii}}$ IRT line profiles in Figure~\ref{Fig1}).

Although two components of V711~Tau are both active stars, the main activity contribution is from the K1~IV primary star. We measure the chromospheric H$_{\alpha}$ emission from both stars as a whole, because the H$_{\alpha}$ emission profile has broad wing and the emission from the G5~V secondary heavily blends with the one from the K1~IV primary. Also, to keep things consistent, the other activity lines are measured in the same way. The equivalent widths (EWs) of the excess emission in the different chromospheric diagnostics are measured on the subtracted spectra using the IRAF/SPLOT task, following the methods described in C15. To analyze the possible rotational modulation of chromospheric activity of V711~Tau, we separate our long-term observations to twelve observing groups and plot the measured EWs of H$_{\alpha}$, H$_{\beta}$, and $\mbox{Ca~{\sc ii}}$ IRT line excess emission as function of orbital phase for each group in Figure~\ref{Fig2}. For each group, Figure~\ref{Fig2b} shows the corresponding correlation between the EWs of the H$_{\alpha}$ line excess emissions and the summed EWs of the $\mbox{Ca~{\sc ii}}$~IRT line excess emissions.

We use the EW$_{8542}$/EW$_{8498}$ ratio as an indicator of the type of chromospheric structure that produces the $\mbox{Ca~{\sc ii}}$ IRT emission. The ratios are obtained around the range of 1.0--2.0, which are consistent with the values in solar plages ($\sim$1.5--3, \citealt{chester1991}) and for several other chromospheric activity stars \citep[e.g.][]{montes2000, gu2002, cao2020}, and therefore suggest that the $\mbox{Ca~{\sc ii}}$~IRT line emission arises from plage-like regions.

Moreover, for the HRS observations, the E$_{H_{\alpha}}$/E$_{H_{\beta}}$ ratios are calculated from the EW(H$_{\alpha}$)/EW(H$_{\beta}$) values with the correction:
\begin{equation}
\frac{E_{H_{\alpha}}}{E_{H_{\beta}}} = \frac{EW(H_{\alpha})}{EW(H_{\beta})}*0.2444*2.512^{(B-R)}
\end{equation}
given by \citet{Hall1992}, which takes into account the absolute flux density in H$_{\alpha}$ and H$_{\beta}$ lines, and the color index ($B-R$~=~1.39 for V711~Tau). We use this ratio as a diagnostic for discriminating the presence of plages or prominences on the stellar surface, following the results of \citet{Hall1992} who found that low E$_{H_{\alpha}}$/E$_{H_{\beta}}$ ($\sim$1--2) can be achieved both in plages and prominences viewed against the disk, but high ratios ($\sim$3--15) can only be achieved in extended regions viewed off the limb. The high ratios (E$_{H_{\alpha}}$/E$_{H_{\beta}}$ $\geq$ 3) that we obtained in V711~Tau indicates that the emission would arise from extended prominence-like regions, in contrast with the plage-like regions inferred from the EW$_{8542}$/EW$_{8498}$ ratios.
\section{Flare characterization}\label{sec4}
\subsection{Flare events}
Solar/stellar flares are powerful and explosive phenomena in the atmosphere, which are commonly thought to be caused by the energy release through the reconnection of magnetic loops and could be detected across the entire electromagnetic spectrum from X-ray to radio wavelength \citep{Schrijver2000}. The flares caused by stars in RS~CVn-type systems are quite often detected, especially in the X-ray, UV, and optical bands.

Due to its very high excitation potential, the $\mbox{He~{\sc i}}$ D$_{3}$ line is a very important indicator to trace out the regions of higher temperature and excitation in solar and stellar chromospheres. When an optical flare event happened, the $\mbox{He~{\sc i}}$ D$_{3}$ line shows obvious emission feature above the continuum level, which has been observed during flares of very active RS CVn-type stars like II~Peg \citep{huenemoerder1987, montes1997, berdyugina1999, frasca2008a}, V711~Tau \citep{garcia2003a, cao2015}, UX~Ari \citep{montes1996, gu2002, cao2017}, DM~UMa \citep{Zhang2016} and SZ~Psc \citep{cao2019, cao2020}. Meanwhile, the other chromospheric activity lines also show stronger emission features during flare events.

There are four optical flare events occurred on V711~Tau during our observations. The first optical flare was observed on 2006 November~29 and 30, and the second one was detected on 2006 December~6. The third flare was detected on 2015 October~30 and 31, and the last one was observed on 2016 January~23. An example of optical flare event (the third one) is displayed in Figure~\ref{Figb3}, in which we plot the observed, synthesized, and subtracted H$_{\alpha}$, $\mbox{Na~{\sc i}}$ D$_{1}$, D$_{2}$ doublet and $\mbox{He~{\sc i}}$ D$_{3}$ line profiles taken in several consecutive observing nights for the pre-flare (2015 October~29), flare state (2015 October~30 and 31), and post-flare (2015 November~1). From Figure~\ref{Figb3}, we can see that the flare diagnostic $\mbox{He~{\sc i}}$ D$_{3}$ line shows weak emission feature on 2015 October~30 and strong emission on October~31 (as indicated by the arrows), and the H$_{\alpha}$ line emissions are stronger during these two nights than the other observations. Correspondingly, stronger excess emission in the $\mbox{Na~{\sc i}}$ D$_{1}$, D$_{2}$ doublet lines could also be seen in the subtracted spectra. According to the measured EWs of the excess H$_{\alpha}$ emission, moreover, the observation at phase 0.182 on 2015 October~31 corresponds to the flare maximum.

V711~Tau is a high-rate flaring star. There are several flare events reported over a wide range of wavelength regions, like X-ray \citep[e.g.][]{garcia2003a, Krimm2006}, UV \citep[e.g.][]{Osten1999}, optical \citep[e.g.][]{foing1994, garcia2003a}, and radio wavelength \citep[e.g.][]{Richards2003}. Moreover, we had also reported two large optical flare events in C15. It is noteworthy that the Swift team detected three bright hard X-ray flares from  V711~Tau. The flares occurred on 2006 November~29 and were captured by the $Burst$ $Alert$ $Telescope$ $(BAT)$ on Swift in 64--second exposures starting at 04:30:39, 04:31:43, and 04:33:51 UTC \citep{Krimm2006}. By comparison, we find that the X-ray flares and our optical flare event detected on 2006 November~29 happened on the same day, and the X-ray flare occurred before the optical one (flare maximum at 17:19:13 UTC). On the one hand, the reason could be that flares may occur at somewhat different times in different wavelength bands. On the other hand, it is possible that we did not capture the whole flare evolution from the initial outburst to the very end due to limited observations. Therefore, we argue that both the X-ray flare-like events and our optical flare may had taken place at the same active region over the surface of V711~Tau. For the other three optical flare events, there are no counterparts reported in any other wavelength bands.

\subsection{Energy released in the H$_{\alpha}$ line}
Based on the the excess EWs, we can estimate the energy released in the chromospheric activity lines during the flares. We compute the stellar continuum flux $F_{H_{\alpha}}$~(erg cm$^{-2}$ s$^{-1}$ \AA$^{-1}$) in the H$_{\alpha}$ line region as a function of the color index $B~-~V$~=~0.92 based on the empirical relationship:
\begin{eqnarray}
\log{F_{H_{\alpha}}}=[7.538-1.081(B-V)]\pm{0.33}\nonumber \\
0.0~\leq~B-V~\leq~1.4
\end{eqnarray}
of \citet{hall1996}, and then convert the EWs into the absolute surface fluxes $F_{S}$~(erg~cm$^{-2}$~s$^{-1}$) through the relation $F_{S}$~=~$F_{H_{\alpha}}$~$\times$~EW. Therefore, the flare energy $L$~(erg~s$^{-1}$) in the observed H$_{\alpha}$ line is derived using the formula $L$~=~4$\pi$R$_{\ast}$$^{2}$$F_{S}$. During the calculation, we assumed that these optical flares occurred on the K1~IV primary star of V711~Tau, and therefore the radius R$_{\ast}$~=~3.9~R$_{\sun}$ is used. Moreover, we have also corrected the EWs to the total continuum before converting to absolute surface fluxes. The absolute surface fluxes and luminosities of these four flare events in the pre-flare, flare maximum, and post-flare periods are listed in Table~\ref{tab2}.

The values for energy released in the H$_{\alpha}$ line during the flares are similar to the ones estimated by us in C15 and have the similar order of magnitude ($\sim$$10^{31}$~erg~s$^{-1}$) to the flares for other RS~CVn-type stars, such as UX~Ari \citep{montes1996, gu2002, cao2017}, HK~Lac \citep{catalano1994} and SZ~Psc \citep{cao2019, cao2020}. For our observations, we had not observed entire flares from the initial outburst to the end, but we can give a rough time scale of 24 hours for our first and third optical flares, respectively. Thus, a total energy emitted in the H$_{\alpha}$ line can be up to the order of magnitude of $10^{36}$~erg for these flares, which are comparable with the values in the other RS CVn flares.

\section{Chromospheric activity variations}\label{sec5}
\subsection{Rotational modulation}
Usually, the distribution of chromospheric activity regions does not uniformly present over the surface of active stars, similar to the hot solar plage regions. Rotational modulation of chromospheric activity has been found in many active stars by means of several chromospheric activity diagnostics \citep[e.g.][]{berdyugina1999, gu2002, garcia2003a, frasca2008a, frasca2008b, zhang2008, cao2014, cao2017, cao2020}.

As shown in Figure~\ref{Fig2}, the ranges of H$_{\alpha}$ line variation are much larger than those of the other lines. And from Figure~\ref{Fig2b}, it can be seen that a positive correlation between the H$_{\alpha}$ and $\mbox{Ca~{\sc ii}}$ IRT line emissions can be clearly recognized, especially for the observing groups of 2008 November, 2015 October--November, 2016 January, 2018 December--2019 January, and 2021 November--December. For the remaining observing groups, a rough positive correlation can still be discerned, although with considerable scatter. Thus, using the H$_{\alpha}$ line excess emission as a representative, we conduct an analysis for the chromospheric active longitudes of V711~Tau. In 2005 November--December observing group, the chromospheric activity level seems much stronger around phases 0.3 and 0.8, which suggests that there are two possible active longitudes over the surface of V711~Tau. In 2006 September--December group, except a strong optical flare event occurred near phase 0.5, there is a clearly modulated trend that two active longitudes could be found around phases 0.25 and 0.75. Also, these active longitudes are much stronger than the previous observing group. However, the chromospheric active longitudes changed and a broader active longitudes seems appeared in 2007 September--December observing group. In 2008 November observing group, a possible stronger active longitude could be found around phase 0.3. During observing groups of 2015 October--November, 2016 January, and 2016 November--2017 January, moreover, it is worth to note that the chromospheric activity variation shows similar behaviors that there are two active longitudes around phases 0.2 and 0.8. As discussed in Section~4.1, an optical flare event took place near phase 0.2 in 2015 October--November observing group, while another flare occurred near phase 0.8 in 2016 January group. To certain extent, these flare events also support our findings of active longitudes. However, from 2017 November to 2020 December groups, the chromospheric active longitudes changed again, and there are no active longitudes over the surface of V711~Tau. The chromospheric activity level is much weaker during these observing groups, which also indicates that there are no obvious active regions on the surface of star. In 2021 November--December observing group, we notice that there is one active longitude located near phase 0.5.

Through the long-term observations, we can identify the distribution of active longitudes on V711~Tau during different observing runs. V711~Tau has two active longitudes near phases 0.25 and 0.75, separating by about 180$\degr$, to dominate the activity most of time. After removing the phase offset, our results are also consistent with the findings in C15.  Moreover, \citet{Jetsu2017} analyzed about 27~years time series differential~V photometric observations from 1987 to 2015, and found that starspots usually concentrate around phases 0.25 and 0.75, which is in good agreement with our results.

\citet{muneer2010} found that the EWs of H$_{\alpha}$ emission of V711~Tau do not show any obvious correlation with the V-band light curve in their analysis. However, we have found a clear anti-correlation between photospheric spots and chromospheric activity of V711~Tau in C15. For our 2005 November--December, 2006 September--December, 2007 September--December, 2008 November, and 2016 November--2017 January observing groups, there are corresponding photometric observations in \citet{muneer2010}, \citet{Jetsu2017} and \citet{Strassmeier2020}. To further investigate the possible spatial correlation between photospheric spots and chromospheric activity in detail, we plot the chromospheric H$_{\alpha}$ variations and phase folded light curves in Figure~\ref{Fig5}. Because of very sparse sampling of the photometric data in 2005 November--December and 2008 November observing groups, we expand the time spans of the data in the comparsion. For other observing groups, the time span is consistent between photometric and chromospheric observations. From Figure~\ref{Fig5}, except the behavior at the second half of the phase in 2006 September--December and 2007 September--December observing groups, it is obvious that there is a good anti-correlation between them at most phases. For example, EWs become stronger when V711~Tau gets fainter, or EWs become weaker when the star gets brighter, which indicates that the chromospheric activity region is associated with the photospheric starspots in spatial structure. It could further concludes that the localized magnetic loop heated the chromospheric activity region is connected to the photospheric spot region.

\begin{figure*}
\centering
\includegraphics[width=15.0cm,height=5.0cm]{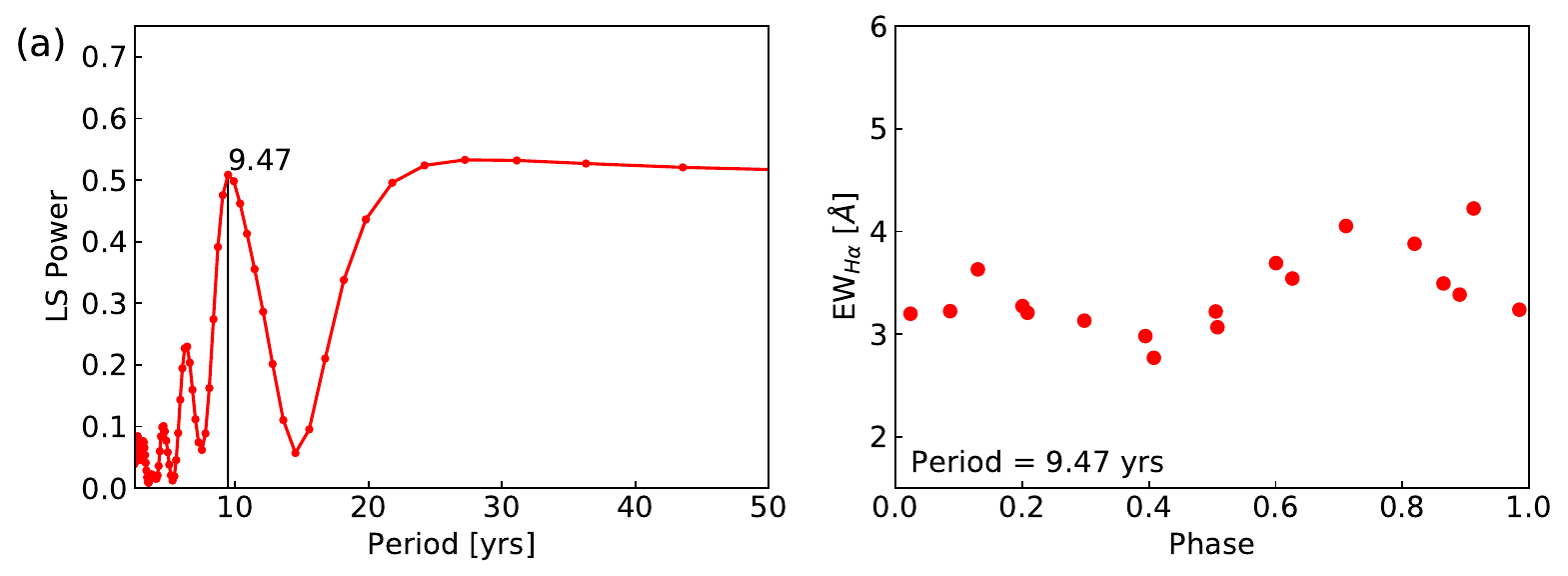}
\includegraphics[width=15.0cm,height=5.0cm]{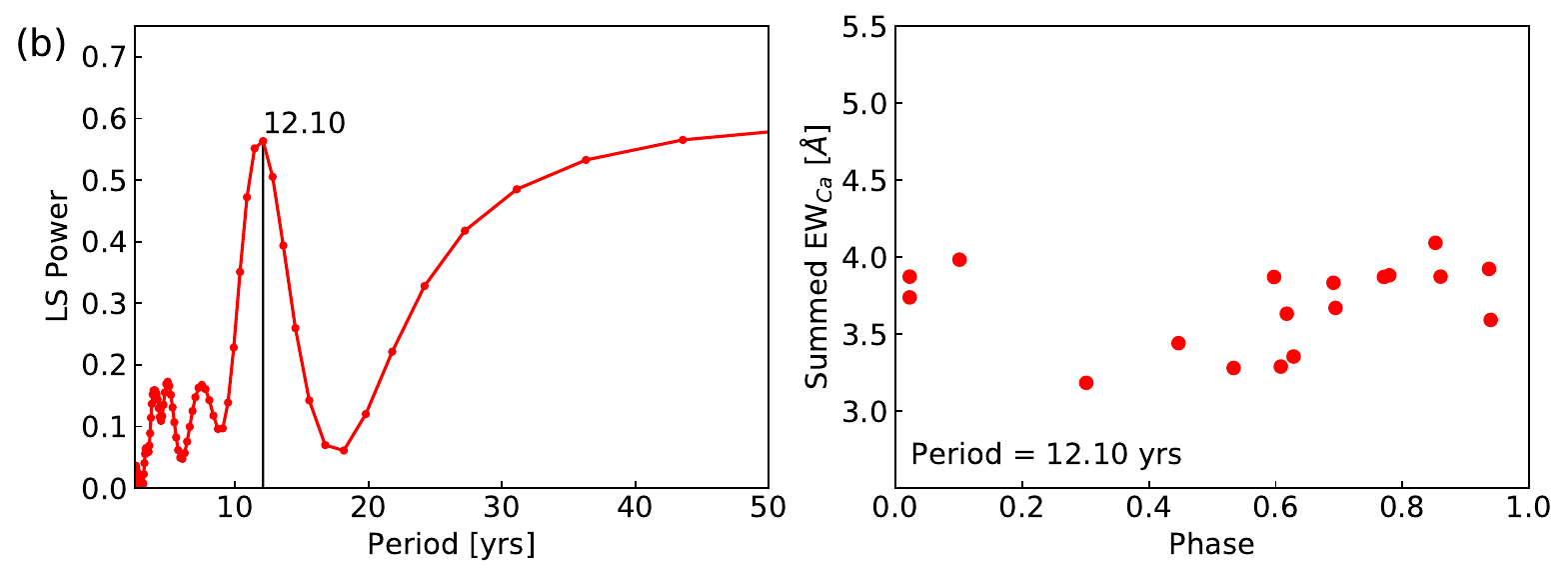}
\caption{(a) Lomb-Scargle periodograms of the H$_{\alpha}$ data (left panel), and phase folded chromospheric activity variations under the assumption of the 9.47~yr (right panel) periods. (b) Lomb-Scargle periodograms of the $\mbox{Ca~{\sc ii}}$ IRT data (left panel) and phase folded chromospheric activity variations under the assumption of the 12.10~yr period (right panel).}
\label{Fig6}
\end{figure*}
\subsection{Long-term variations}
The EWs of the H$_{\alpha}$ excess emission, along with the summed EWs of the $\mbox{Ca~{\sc ii}}$ IRT excess emissions, are plotted against the observing time in the top and middle panels of Figure~\ref{Fig4}, where the data from C15, \citet{montes1995a}, and \citet{montes1997} are also included. Here we have corrected all the EWs to the total continuum, taking into account of the different intensity weight ratios adopted in the synthesized spectra construction due to different reference stars used. For our observations, it can be seen that the data points exhibit significant dispersion in some observing groups due to rotational modulations and the sampling is fragmented. Therefore, to effectively search for any cyclic behavior, it is essential to minimize these influences. For each observing group with better phase coverage, we first exclude the flare data points and subsequently fit a linear function to represent the trend of the chromospheric activity. Then, an averaged data point is derived by calculating the mean between the maximum and minimum values obtained from the linear fitting. We use its value to represent the overall activity level of V711~Tau in this observing group, which is illustrated as a red dot in Figure~\ref{Fig4}.

It can be seen that the chromospheric activity level starts to get stronger since the minimum near 2002 and reaches to a maximum around the year of 2006. And then the activity level shows a change trend from high to low, especially for the H$_{\alpha}$ line, although there is an unfortunate gap between 2008 and 2015. A long-term light curve of V711~Tau is also displayed in the bottom panel of Figure~\ref{Fig4}, to show the comparsion between the chromospheric activity variations and the behavior of photospheric starspots in long-term scale. During the years from 1986 to 1995, probably because of sparse sampling of chromospheric activity data, the activity variation has no clearly relevance to the photometric light curves. However, it is obvious that there is a good anti-correlation during the years from 1998 to 2008, which also suggests that there is a close spatial connection between photospheric starspots and chromospheric active regions in long time scale.

The cyclic behavior with 19.5~$\pm$~2.0~yrs in photospheric starspot activity of V711~Tau had been derived by \citet{lanza2006}, 15~--~16~yrs by \citet{berdyugina2007}, 14.1~$\pm$~0.3 yrs by \citet{muneer2010}, and 14.8~yrs by \citet{Perdelwitz2018}. Also, possible chromospheric activity cycle with a period of about 18 years had been derived by \citet{buccino2009}, based on the mean annual Mount Wilson index $S$ from the $IUE$ spectra. Recently, \citet{Perdelwitz2018} tried to independently determine the activity cycle of V711~Tau using archival X-ray data, but failed because of sparse sampling of data and stochastic flaring activity. To search for the possible long-term periodic chromospheric variations, we use the Lomb-Scargle periodogram to analyze the averaged H$_{\alpha}$ and $\mbox{Ca~{\sc ii}}$~IRT data (the red dots in Figure~\ref{Fig4}), respectively. The H$_{\alpha}$ data points before the year of 2000 are excluded from the analysis due to the sparse coverage. The resulting periodograms are displayed in Figure~\ref{Fig6}. In the case of the H$_{\alpha}$ line, there is a possible period of about 9.47~yrs with a False Alarm Probability (FAP) of about 4\%. The corresponding phase folded chromospheric H$_{\alpha}$ variations is also shown in Figure~\ref{Fig6}. For the $\mbox{Ca~{\sc ii}}$ IRT lines, we also derived a possible period of about 12.10~yrs with a FAP of about 2\% (also see Figure~\ref{Fig6}). The period is close to those derived by \citet{muneer2010} and \citet{Perdelwitz2018} for photospheric starspot activity. Although there is an observing gap between the year of 2008 and 2015 for our situation, the similar activity cycles can be speculated from the data between 2000 and 2008 presented in Figure~\ref{Fig4}. The $\mbox{Ca~{\sc ii}}$ IRT lines are anticipated to provide a more accurate representation of plage-like regions on the stellar surface, whereas the H$_{\alpha}$ line appears to exhibit an additional contribution from above-limb sources. Therefore, it is possible that the activity cycles of the H$_{\alpha}$ and $\mbox{Ca~{\sc ii}}$ IRT lines exhibit distinct characteristics. The result from the $\mbox{Ca~{\sc ii}}$ IRT lines suggests that chromospheric plage-like regions exhibit similar long-term pattern as photospheric starspots for V711~Tau.

\section{Summary and conclusions}\label{sec6}
In this paper, we have investigated long-term high-resolution spectroscopic observations of V711~Tau taken during several observing runs from 2005 to 2021. Through the study of optical flares, rotational modulation and long-term variation of chromospheric activity, as well as the correlation between photospheric and chromospheric activity, we draw the following main conclusions and the new results:
\begin{enumerate}
\item V711~Tau shows the excess emission features in the $\mbox{Ca~{\sc ii}}$ IRT, H$_{\alpha}$, and H$_{\beta}$ lines, and the chromospheric emission is mainly associated with the K1~IV primary star. The G5~V secondary star also has active chromosphere. These results are consistent with previous findings.

\item There are four optical flares detected during our long-term observations. Flare energies, released in the chromospheric H$_{\alpha}$ line during flare maximum, have similar values to the ones for V711~Tau estimated by us in C15 and have the similar order of magnitude to the flares of other RS CVn-type stars.

\item To analyze the possible rotational modulation of chromospheric activity of V711~Tau, we divide our long-term observations to several observing groups. We find that V711~Tau has two active longitudes around phases 0.25 and 0.75 to dominate the activity most of time, which separate by about 180$\degr$. 

\item The chromospheric activity level of V711~Tau shows a long-term variation. Possible chromospheric activity cycles are derived as 9.47~yrs for the H$_{\alpha}$ line and 12.10~yrs for the $\mbox{Ca~{\sc ii}}$ IRT lines. Moreover, a close spatial connection between photospheric spots and chromospheric active regions in short and long time scale can be found, which indicate the active regions at different atmospheric layers should have the same origin.
\end{enumerate}

\begin{acknowledgments}
The authors thank the anonymous referee for the careful review and helpful suggestions, which lead to a significant improvement in our manuscript. The authors acknowledge the support of the staff of the Xinglong 2.16m telescope. This work was partially Supported by the Open Project Program of the CAS Key Laboratory of Optical Astronomy, National Astronomical Observatories, Chinese Academy of Sciences. The authors would like to thank the support of the staff of the Lijiang 2.4m telescope. Funding for the telescope has been provided by Chinese Academy of Sciences and the People's Government of Yunnan Province. The present study is financially supported by the National Natural Science Foundation of China (NSFC) under grants Nos. 10373023, 10773027, U1531121, and 11903074, and also supported by the Yunnan Fundamental Research Projects (grant Nos. 202201AT070186 and 202305AS350009), the Yunnan Revitalization Talent Support Program (Young Talent Project), and International Centre of Supernovae, Yunnan Key Laboratory (No. 202302AN360001). The authors also acknowledge the science research grant from the China Manned Space Project.
\end{acknowledgments}
\bibliography{sample631}{}
\bibliographystyle{aasjournal}

\end{document}